\documentclass[twocolumn,showpacs,prb]{revtex4}
\usepackage{amsmath}
\usepackage{amssymb}
\usepackage{graphicx}

\begin{document}

\title{Non-local detection of resistance fluctuations of an open quantum dot}

\author{A. I. Lerescu}
\affiliation{Physics of Nanodevices Group, Zernike Institute for Advanced Materials,\\
University of Groningen, Nijenborgh 4, 9747 AG  Groningen, The Netherlands}
\author{J. H. Bardarson}
\affiliation{Instituut-Lorentz, Universiteit Leiden, P.O. Box 9506,
2300 RA Leiden, The Netherlands}
\author{E. J. Koop}
\affiliation{Physics of Nanodevices Group, Zernike Institute for Advanced Materials,\\
University of Groningen, Nijenborgh 4, 9747 AG  Groningen, The
Netherlands}
\author{C. H. van der Wal}
\affiliation{Physics of Nanodevices Group, Zernike Institute for Advanced Materials,\\
University of Groningen, Nijenborgh 4, 9747 AG  Groningen, The
Netherlands}
\author{B. J. van Wees}
\affiliation{Physics of Nanodevices Group, Zernike Institute for Advanced Materials,\\
University of Groningen, Nijenborgh 4, 9747 AG  Groningen, The
Netherlands}

\date{\today}

\begin{abstract}
We investigate quantum fluctuations in the non-local resistance of
an open quantum dot which is connected to four reservoirs via
quantum point contacts. In this four-terminal quantum dot the
voltage path can be separated from the current path. We measured
non-local resistance fluctuations of several hundreds of Ohms, which
have been characterized as a function of bias voltage, gate voltage
and perpendicular magnetic field. The amplitude of the resistance
fluctuations is strongly reduced when the coupling between the
voltage probes and the dot is enhanced. Along with experimental
results, we present a theoretical analysis based on the
Landauer-B\"{u}ttiker formalism. While this theory predicts
non-local resistance fluctuations of about 20 times larger amplitude
than what has been observed, agreement with theory is very good if
it is scaled with a factor that accounts for the influence of
orbital dephasing inside the dot. This latter case is in reasonable
agreement with an independently determined time scale for orbital
dephasing in the dot.
\end{abstract}

\pacs{
73.23.-b,       
73.63.Kv,       
73.40.-c}       


\maketitle

\section{\label{sec:introduction}Introduction}
When a conducting solid-state system is smaller than the phase
coherence length of the electrons, its electrical conductance is
significantly influenced by quantum interference. For diffusive thin
films this results in phenomena known as universal conductance
fluctuations and weak localization
\cite{altshuler1998phystod,beenakker1991ssp,beenakker1997rmp,imry2002book}.
Similar conductance fluctuations and localization phenomena are
observed in micron-scale ballistic quantum dots, since these behave
in practice as chaotic cavities due to small shape irregularities in
the potential that defines the dot. These conductance fluctuations
have been extensively studied for two-terminal quantum dots
\cite{huibers1998Aprl,huibers1998Bprl,alves2002prl,zumbuhl2005prb},
\textit{i.e.}\ systems with only a source and a drain contact.
However, for quantum dots this two-terminal conductance is often
influenced by Coulomb blockade and weak localization effects, which
complicate an analysis when one aims at studying other effects.

\begin{figure}[p]
\includegraphics[width=8cm]{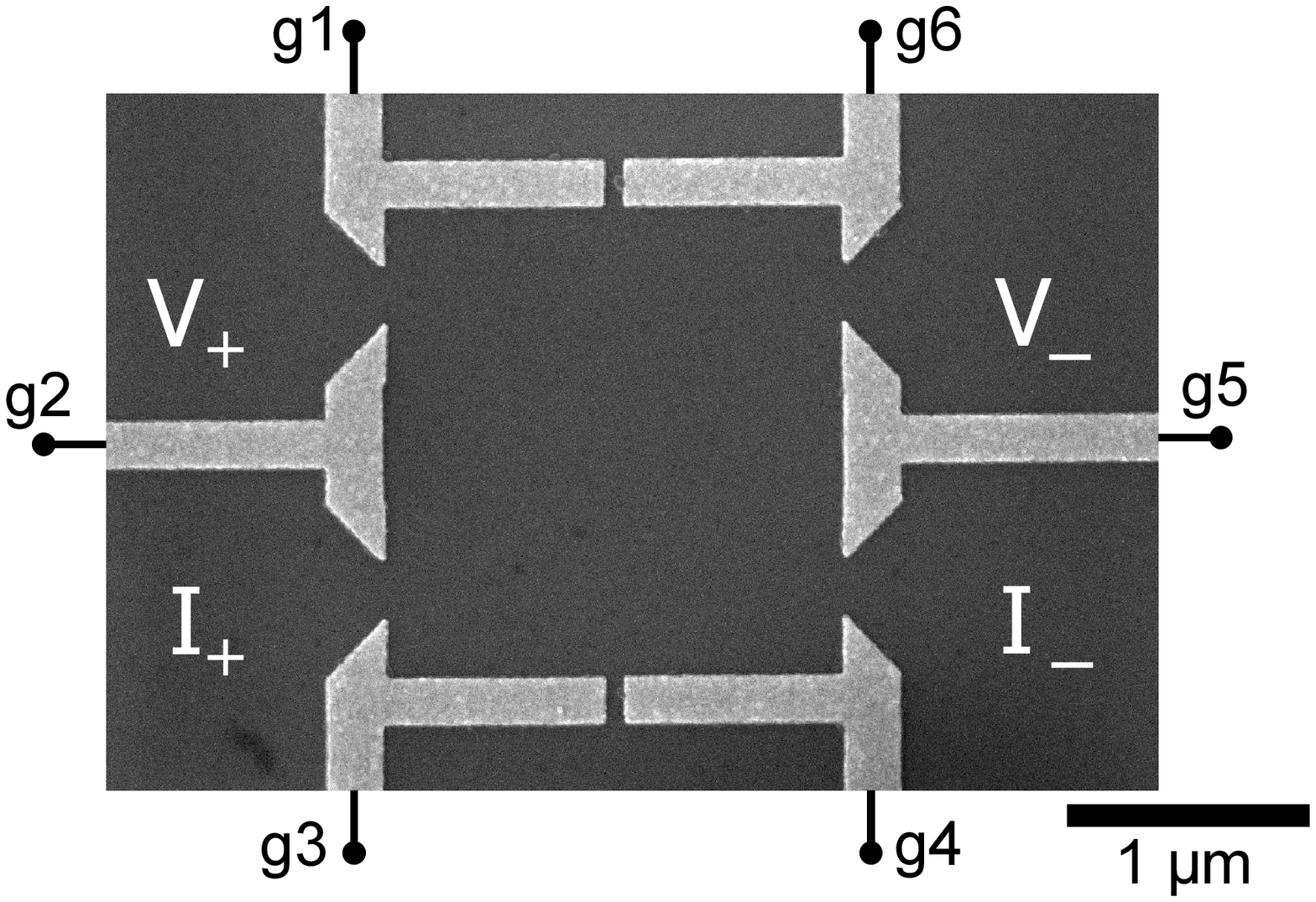}
\caption{\label{fig:device}
Electron microscope image of the device studied in this article.
The position of the reservoirs used for current biasing
($I_{+}$ and $I_{-}$) and voltage probes ($V_{+}$ and $V_{-}$)
is indicated, as well as the numbering of the gates labeled $g1$-$g6$.
Unless stated otherwise, all results presented in this article were obtained in this
non-local configuration.}
\end{figure}

We present here a study of fluctuations in electron transport in a
\textit{four}-terminal ballistic quantum dot. The dot is coupled to
four reservoirs via quantum point contacts (QPC). In such a system,
the voltage path (with probes at voltage $V_{+}$ and $V_{-}$) can be
separated from the path that is used for applying a bias current $I$
(see Fig.~\ref{fig:device}). Consequently, one can measure so-called
\textit{non-local}
\cite{buttiker1986prl,benoit1987prl,skocpol1987prl,haucke1990prb}
voltage signals that are purely due to quantum fluctuations of the
chemical potential \cite{buttiker1989prb} inside the dot, and for
which a naive classical analysis predicts a signal very close to
zero. For linear response, this is expressed as a non-local
resistance $R_{nl}=(V_{+}-V_{-})/I$ (this non-local resistance will
fluctuate around a value that is very close to zero Ohm, and is
therefore studied in terms of resistance rather than conductance).
Increasing the number of open channels in the voltage probes will
result in enhanced dephasing for the electronic interference
effects. With a four-terminal system, one can study this directly
since it results in a reduction of the amplitude of the non-local
resistance fluctuations. Notably, such a reduction of the
fluctuation amplitude does not occur upon increasing the number of
open channels in a two-terminal system \cite{baranger1994prl}.
Furthermore, such a four-terminal systems could be used for studying
signals that are due to spin. In a strong magnetic field QPCs can be
operated as spin-selective injectors or voltage probes
\cite{potok2002prl}. This can be used to generate and detect an
imbalance in the chemical potential for spin-up and spin-down
electrons \cite{folk2007aps}, similar to non-local spin-valve
effects observed in metallic nanodevices \cite{jedema2001nature}.
Also here, a four-terminal dot is an interesting alternative to work
on spin physics in dots with two-terminal devices
\cite{folk2003science,zumbuhl2001aps,beenakker2006prb}. However, if
such a system is smaller than the electron phase coherence length,
the non-local signals with information about spin will also show
fluctuations that result from interference of electron trajectories
\cite{bardarson2006condmat}.

In this article, we focus on our first experiments with such a
four-terminal quantum dot. We aimed at characterizing the non-local
resistance fluctuations, and studying the influence of the voltage
probes on the typical amplitude of these fluctuations. As a
comparison with our experimental results we present a numerical
simulation of the non-local resistance, based on the
Landauer-B\"{u}ttiker formalism
\cite{buttiker1986prl,landauer1957ibm} and the kicked rotator
\cite{izrailev1990physrep,fyodorov2000jetp}.

The outline of the paper is as follows:
Section~\ref{sec:experimental} presents the experimental
realization. In section~\ref{sec:resflucdata}, we present
measurements of the non-local resistance as a function of bias
voltage, gate voltage, and magnetic field, and confirm that the
observed fluctuations in the non-local resistance are the
four-terminal equivalent of universal conductance fluctuations in
two-terminal systems. In section~\ref{sec:probecoupling}, we analyze
how the typical amplitude of the measured non-local resistance
fluctuations depends on the number of open channels in the voltage
probes. Section~\ref{sec:theory} presents our theoretical analysis
with a comparison to the experimental results, before ending with
conclusions.

\section{\label{sec:experimental}Experimental realization}

Our device was fabricated using a ${\rm GaAs}/{\rm Al}_{x}{\rm
Ga}_{1-x}{\rm As}$ heterostructure containing a two dimensional
electron gas (2DEG) 75 nm below the surface, purchased from Sumitomo
Electric Co. At 4.2 K, the mobility was $\mu = 86 \; {\rm
m}^{2}/{\rm Vs}$ and the electron density was $n_{s} = 2.4 \cdot
10^{15} \; {\rm m}^{-2}$. The dot was designed with an area of $2
\times 2 \; \mu {\rm m}^{2}$. Figure \ref{fig:device} shows and
electron microscope image of the device. Six depletion gates were
deposited on the surface (15 nm of Au with a Ti sticking layer) and
were used for defining the dot in the 2DEG. We estimate that the
depletion width around the gates was about $100 \; \rm{nm}$, such
that the electron gas area $A_{dot}$ inside the dot was about $3.2
\; \rm{ \mu m ^2 }$. With these six gates the dot could be coupled
to the four reservoirs via QPCs in a controllable manner. All four
QPCs showed clear quantized conductance steps
\cite{vanwees1988prl,wharam1988jpc} in measurements where only the
corresponding pair of gates were depleting the 2DEG. Note that
throughout this article we use that a QPC with a conductance of
$2e^{2}/h$ is defined as having one open channel (denoted as $N=1$),
\textit{i.e.}\ we neglect spin when counting channels. The four
reservoirs were connected to macroscopic leads via Ohmic contacts,
which were realized by annealing a thin Au/Ge/Ni layer that was
deposited on the surface.

All the measurements were performed with the sample at a temperature
of 130~mK. However, the temperature dependence of our data saturated
when cooling below $\approx 400 \; \rm{mK}$, so we will assume this
value for the effective electron temperature. We used a current bias
$I$ with standard ac lock-in techniques at 13 Hz. Unless stated
otherwise, we used $I=1 \; {\rm nA}$. The non-local resistance
$R_{nl}$ was then recorded as the zero-bias differential resistance
$dV/dI$, with $V$ defined as $V \equiv V_{+}-V_{-}$. We used a
floating voltmeter to measure $V$, thus being insensitive to the
voltage across the dot along the current path, and thereby
insensitive to Coulomb blockade and weak localization effects. On
the current path, only the $I_{-}$ reservoir was connected to the
grounded shielding of our setup, and all gate voltages were applied
with respect to this ground.

A magnetic field could be applied, with
an angle of 7$^{\mathrm{o}}$ with respect to the 2DEG plane
(determined from standard Hall measurements and electron focusing
effects, discussed below). The perpendicular component of this field
was used for studying the dependence of the non-local resistance on
perpendicular magnetic field. The component of the magnetic field
parallel with the 2DEG plane was oriented perpendicular to the
current path. While this parallel field was about ten times stronger
than the perpendicular field, the orbital effects associated with
this parallel field are negligibly small, and it can be disregarded
for all of the experimental results presented here (and weak enough
to not significantly reduce the amplitude of resistance fluctuations
\cite{debray1989prl,folk2001prl,zumbuhl2004prb}).

\section{\label{sec:resflucdata}Non-local resistance fluctuations}

Figure~\ref{fig:gatedependence} shows the non-local resistance
$R_{nl}$ as a function of the voltage applied to gate $g1$ and gate
$g6$. The other four gate voltages were kept constant during this
measurement, with the QPCs in the current path at a conductance of
$2e^{2}/h$ each (one open channel, $N=1$). The range of gate voltage
for $g1$ and $g6$ used here corresponds to opening the voltage-probe
QPCs from nearly pinched off ($N=0$) up to about $N=8$ open
channels. As a function of these gate voltages, the non-local
resistance shows a random pattern of fluctuations around zero Ohm,
with maximums and minimums up to about $\pm 500 \; \Omega$. Notably,
the change in gate voltage needed to change $R_{nl}$ significantly
(one fluctuation), is very similar to the change in gate voltage
needed for increasing the number of open channels in a QPC by one.
This corresponds to changing the shape of the potential that forms
the dot by a distance of about half a Fermi wavelength, which is
consistent with the length scale needed for significantly changing a
random interference pattern of electron trajectories. These
non-local resistance fluctuations as a function of the gate voltage
on $g1$ and $g6$ were highly reproducible, and indeed a so-called
finger print of the sample. The identical measurement repeated after
4 days (during which we performed strong magnetic field sweeps and a
temperature cycle up to 4.2 K) showed nominally the same fluctuation
pattern as in Fig.~\ref{fig:gatedependence}.

In Fig.~\ref{fig:thouless} we present results of studying the
dependence of the amplitude of the non-local resistance fluctuations
on the amplitude of the applied bias current $I$. The figure shows
measurements of $R_{nl}$ as a function of the gate voltage $V_{g1}$.
The results show several fluctuations that are reproducible, but
decreasing in amplitude upon increasing the amplitude of the bias
current. In this experiment, the conductance of the other three QPCs
was fixed at $2e^{2}/h$. The amplitude of the fluctuations reduces
when the measurement averages over contributions of electrons in
uncorrelated orbitals, that is, averaging over electrons that differ
in energy by more than the Thouless energy
\cite{beenakker1997rmp,imry2002book}. When the current bias is
increased, the corresponding voltage bias $V_{bias}$ increases as
well (see labels in the Fig.~\ref{fig:thouless}), and this is used
to experimentally estimate the Thouless energy $E_{Th}$ for our
system. The amplitude of the fluctuations starts to decrease
significantly around $V_{bias} \approx 125 \; {\rm \mu V}$. This is
close to a theoretical estimate
\cite{marcus1997chasolfrac,huibers1999prl} for $E_{Th} = \frac{\hbar
v_{F}}{L} \approx 80 \; {\rm \mu eV}$, where $v_{F}$ the Fermi
velocity and $L$ the effective width of the dot.

We now turn to measurements of the non-local resistance as a
function of perpendicular magnetic field, presented in
Fig.~\ref{fig:bfield}a. Here the conductance of all four QPCs was
fixed at $2e^{2}/h$. The trace of $R_{nl}$ shows random fluctuations
of similar amplitude as observed in the gate voltage dependence. For
estimating the typical magnetic field scale that significantly
changes the value of $R_{nl}$ (the correlation field $\Delta
B_{c}$), we apply an established method from such studies on similar
two-terminal quantum dot systems (following
Refs.~\onlinecite{marcus1993prb,chan1995prl}). For this, we took the
averaged power spectrum $S_B(f_B)$ of traces as in
Fig.~\ref{fig:bfield}a (only using the low-field data in the range
$\pm 140 \; {\rm mT}$, see below). On a logarithmic scale,
$S_B(f_B)$ closely resembles a straight line with negative slope in
the frequency range between $f_B = 0.05$ and 0.5 cycles per mT (and
then levels off), very similar to the results from the studies with
two-terminal dots \cite{marcus1993prb,chan1995prl}. We fit this part
of the spectrum to the form predicted by semiclassical theory
\cite{marcus1993prb,chan1995prl}, $S_B(f_B) = S_B(0) (1 + 2\pi
\alpha \phi_0 f_B) \cdot \exp{(- 2\pi \alpha \phi_0 f_B} )$, where
$\phi_0 = h/e$ the flux quantum, and $\alpha$ the inverse of an
effective area for orbitals in the dot which defines $\Delta B_{c} =
\alpha \phi_0$. This yields $\Delta B_c = 2.1 \pm 0.8 \rm{mT}$ (the
large error bar must be assumed since we could only measure a small
number of independent fluctuations for this analysis). This is in
good agreement with theory for universal conductance fluctuations
\cite{beenakker1997rmp}, which predicts $\Delta B_{c} \approx
\frac{\phi_0}{A_{dot}} = 1.3 \; {\rm mT}$ (simply expressing the
magnetic field needed for adding one flux quantum $\phi_0$ through
the area of the dot). It is commonly observed that $\Delta B_{c}$ is
enhanced by a factor up to about $\sim 2$ due to flux cancelation
effects for electrons that move ballistically between the edges of
the dot
\cite{beenakker1988prb,beenakker1991ssp,marcus1997chasolfrac}. The
measured value for $\Delta B_{c}$ is also in agreement with the
observation of a weak-localization peak around zero magnetic field
\cite{beenakker1997rmp} observed in the two-terminal resistance
(breaking the time-reversal symmetry), which has a width of the same
order of magnitude as $\Delta B_{c}$.

For confirming that changing the field by more than $\Delta B_{c}$
gives access to a statistically independent set of fluctuations, we
studied fluctuations in $R_{nl}$ as a function of gate voltage
$V_{g1}$, for different values of the perpendicular field
(Fig.~\ref{fig:bfield}b). This data confirms that changing the
perpendicular magnetic field in steps of 14~mT gives access to
completely different patterns of random fluctuations in $R_{nl}$.

The inset of Fig.~\ref{fig:bfield}a shows the appearance of much
higher peaks in $R_{nl}$ (up to $3 \; {\rm k} \Omega$) for
perpendicular magnetic fields stronger than $\pm 140 \; {\rm mT}$.
We could confirm that these peaks are due to electron focusing and
skipping orbit effects. With only the three gates $g1$, $g2$ and
$g3$ depleting the 2DEG, our device is identical to devices used for
electron focusing experiments by Van Houten {\it et al.}
\cite{vanhouten1989prb}, and we observe very similar focusing peaks
as in this work at only one polarity of the magnetic field. With the
dot formed, these effects cause peaks in $R_{nl}$ at both polarities
of the magnetic field. The onset of these effects at $\pm 140 \;
{\rm mT}$ agrees with a focusing radius of about $1 \; \mu {\rm m}$.

\section{\label{sec:probecoupling}Influence of voltage probes}

\begin{figure}[p]
\includegraphics[width=8cm]{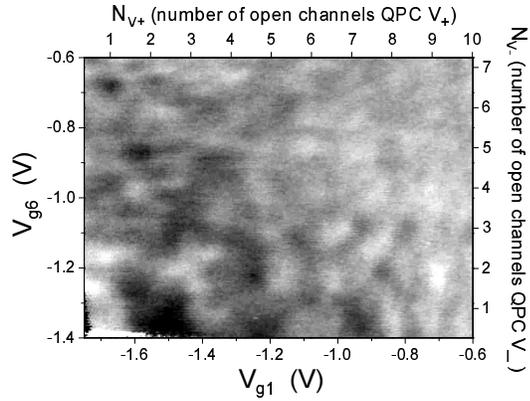}
\caption{\label{fig:gatedependence} Gray scale plot of the non-local
resistance $R_{nl}$ as a function the voltages applied to gates $g1$ and $g6$.
The axes also show the corresponding number of open
channels for the $V_{+}$ and $V_{-}$ probes.
The gray scale shows $R_{nl}$ at a scale from $-500 \; \Omega$ (black)
to $500 \; \Omega$ (white).
The value of $R_{nl}$ fluctuates around zero Ohm, with a typical amplitude that decreases
when the number of open channels in the $V_{+}$ and $V_{-}$ probes
increases. The QPCs formed by gates $g2$ and
$g3$, as well as $g4$ and $g5$ (defining the current path) had a fixed
conductance of $2e^{2}/h$ each. Data taken in zero magnetic field at 130 mK.}
\end{figure}

\begin{figure}[p]
\includegraphics[width=8cm]{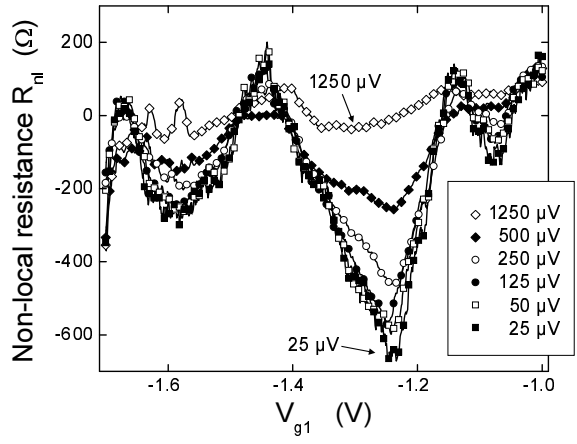}
\caption{\label{fig:thouless} The non-local resistance
$R_{nl}$ as a function of the
voltage $V_{g1}$ applied to gate $g1$, taken for different amplitudes of
the bias current $I$ in the lock-in detection scheme (from 1 nA up to 50 nA).
The legend shows the corresponding values for the voltage drop across
the quantum dot along the current path, obtained as
$V_{bias} \approx I \times \frac{2}{2e^{2}/h}$. The curves show a
reduction of the amplitude of
the non-local resistance fluctuations with increasing $V_{bias}$.
The conductance of the QPC formed by $g5$ and $g6$
(defining the $V_{-}$ probe) is set at $2e^{2}/h$.
For further experimental parameters see Fig.~\ref{fig:gatedependence}.}
\end{figure}

\begin{figure}[p]
\includegraphics[width=8cm]{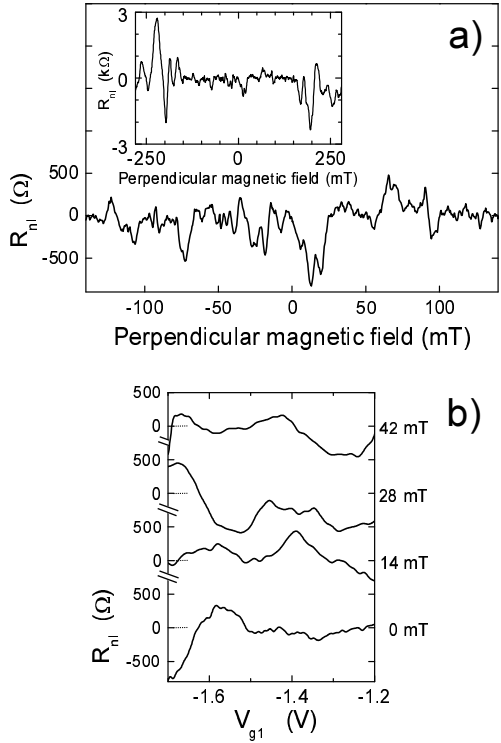}
\caption{\label{fig:bfield} a) The non-local resistance $R_{nl}$
as a function of magnetic field.
The magnetic field is given on the scale of the
perpendicular component of the total applied field.
For this measurement all four QPCs are defined to have a conductance of $2e^{2}/h$.
The inset presents the same data for a wider range of the magnetic field, showing
the onset of electron focusing effects for perpendicular fields larger than $\pm$ 140 mT.
The curves in b) show the non-local resistance as a function of the
voltage applied to gate $g1$ at different values of the perpendicular magnetic field.
The conductance of the QPC formed by $g5$ and $g6$ (defining the $V_{-}$ probe) is kept at $2e^{2}/h$.
For further experimental parameters see Fig.~\ref{fig:gatedependence}.}
\end{figure}

The presence of additional voltage probes on a quantum dot
system will act as source of dephasing for the electrons in
the dot, and this effect should increase when the coupling
between the dot and the probe reservoirs is enhanced.
Earlier work recognized that non-local voltage probes
on a mesoscopic system are a
source of dephasing \cite{kobayashi2002jpsjap,seelig2003prb},
and in theoretical work an additional voltage
probe is often used to model a source of dephasing
\cite{brouwer1995prb,brouwer1997prb}.
This can be directly studied with our system.
The amplitude of the non-local resistance fluctuations
(which result from electron phase coherence) should decrease
when the voltage probes are tuned to carry more open channels.
To study this effect we used data sets of the type presented in
Fig.~\ref{fig:gatedependence}. We concentrate on the case where
the time-reversal symmetry is broken ($\beta = 2$) by applying
weak magnetic fields, since this allows us to get
statistics from a larger set of data.

For a data set as in Fig.~\ref{fig:gatedependence}, the total number
of open channels in the voltage probes ($N_{V+} + N_{V-}$) is lowest
in the bottom left corner of the graph, and highest in the top right
corner. Inspection of $R_{nl}$ in Fig.~\ref{fig:gatedependence}
confirms that the typical amplitude of the fluctuations decreases
when the voltage probes get more open channels. For a more
quantitative analysis of this observation, we determined the mean $
\langle R_{nl} \rangle $ and root-mean-square (rms) standard
deviation $\Delta R_{nl}$ of the non-local resistance for traces
recorded at a fixed number of channels in the voltage probes. This
can be obtained by following $R_{nl}$ along lines with constant
$N_{V+}$ + $N_{V-}$. The theory in the next section shows that, on
such a line, $\Delta R_{nl}$ should also show a weak dependence on
$N_{V+}$ - $N_{V-}$. However, we do not have sufficient data to
study this, and simply average along lines with constant $N_{V+}$ +
$N_{V-}$. The results of this analysis are presented in
Fig.~\ref{fig:flucanalysis}. The large error bars for $ \langle
R_{nl} \rangle $ and $\Delta R_{nl}$ in Fig.~\ref{fig:flucanalysis}
are due to the fact that we could only record a finite number of
independent data sets with fluctuations in $R_{nl}$ (see
Ref.~\onlinecite{errorbars} for further details).

The results in Fig.~\ref{fig:flucanalysis} confirm that $ \langle
R_{nl} \rangle $ is very close to zero, for all values of $N_{V+} +
N_{V-}$. More interestingly, $\Delta R_{nl}$ smoothly decreases as a
function $N_{V+} + N_{V-}$, demonstrating directly the dephasing
influence of the voltage probes for the electrons in the quantum
dot. The typical fluctuation amplitude approaches zero when the dot
becomes fully open (very strong coupling to a reservoir). For a
quantitative evaluation of this observation, we will first present a
theoretical model in the next Section.

\section{\label{sec:theory}Theoretical analysis and discussion}

For our theoretical modeling we consider a ballistic chaotic cavity
connected to four reservoirs through quantum point contacts. A net current $I$
flows between two of the contacts (from $I_{+}$ to $I_{-}$),
while there is no net current flowing into two contacts used as voltage probes
(contacts $V_{+}$ and $V_{-}$).
We use the Landauer-B\"{u}ttiker formalism to derive
the relations between the current $I$ and the voltages
of the four contacts \cite{bardarson2006condmat},
\begin{widetext}
\begin{subequations}
\begin{equation}\label{eq1a}
  I = \frac{1}{2}  \left( \frac{2e^{2}}{h} \right) \lbrack
   \left( (N_{1}-T_{11})+(N_{2}-T_{22})+T_{12}+T_{21} \right) \frac{V_{bias}}{2}+(T_{23}-T_{13})V_{3}+(T_{24}-T_{14})V_{4}
  \rbrack ,
\end{equation}
\begin{equation}\label{eq1b}
  V_{3} = \frac{V_{bias}}{2} \frac{(N_{4}-T_{44})(T_{31}-T_{32})+T_{34}(T_{41}-T_{42})}{(N_{3}-T_{33})(N_{4}-T_{44})-T_{34}T_{43}} ,
\end{equation}
\begin{equation}\label{eq1c}
  V_{4} = \frac{V_{bias}}{2} \frac{(N_{3}-T_{33})(T_{41}-T_{42})+T_{43}(T_{31}-T_{32})}{(N_{3}-T_{33})(N_{4}-T_{44})-T_{34}T_{43}} ,
\end{equation}
\end{subequations}
\end{widetext}
where we used (for concise labeling) notation according to
\[
\begin{array}{c}
I_{+} \; \; \leftrightarrow \; \; 1 , \\
I_{-} \; \; \leftrightarrow \; \; 2 ,\\
V_{+} \; \; \leftrightarrow \; \; 3 ,\\
V_{-} \; \; \leftrightarrow \; \; 4 .
\end{array}
\]
Here the $T_{ij}$ are the transmission probabilities from contact
$i$ to $j$, while the $N_{i}$ are the number of open channels in
contact $i$. The voltages $V_{i}$ are all defined with respect to a
ground \cite{groundingnote} which is defined such that $V_{1} =
+V_{bias}/2$ and $V_{2} = -V_{bias}/2$, where $V_{bias}=V_{1}-V_{2}$
is the voltage across the dot in the current path that is consistent
with a bias current $I$. The measured voltage $V$ in the experiments
corresponds to the quantity $V = V_{3} - V_{4}$, and the non-local
resistance is then
\begin{equation}
\label{Rnltheory}
R_{nl} = \frac{V_{3} - V_{4}}{I} .
\end{equation}

We obtain the mean and root-mean-square (rms) of the $R_{nl}$ by
generating a set of random scattering matrices with the kicked
rotator \cite{izrailev1990physrep,fyodorov2000jetp}. The kicked
rotator gives a stroboscopic description of the dynamics in the
quantum dot, which is a good approximation of the real dynamics for
time scales larger then the time of flight across the dot. The
particular implementation we used is described in detail in Ref.\
\onlinecite{bardarson2005prb}. In a certain parameter range, this
model gives results which are equivalent to random matrix theory
\cite{beenakker1997rmp}. In our simulations we use parameters in
this range, the details of which can be found in Ref.\
\onlinecite{bardarson2005prb}.

\begin{figure}[p]
\includegraphics[width=8cm]{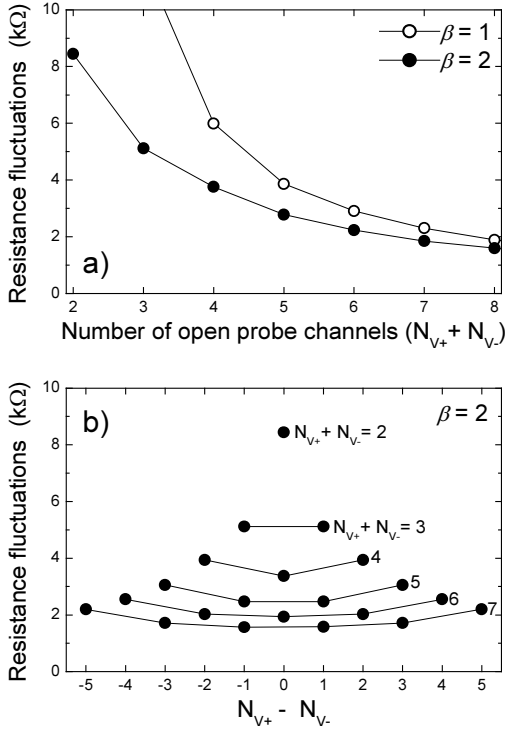}
\caption{\label{fig:voltfluc} a) Theoretical rms values of fluctuations in the non-local resistance ($\Delta R_{nl}$),
as a function of the total number open of channels in the voltage probes $N_{V+}+N_{V-}$.
The two curves present results for a system with ($\beta=1$) and without ($\beta=2$) time-reversal symmetry.
The QPCs for the current path were assumed to have a conductance of $2e^{2}/h$.
The data is obtained from a Landauer-B\"{u}ttiker description of the quantum dot
system and numerical simulations based on random matrix theory.
b) Theoretical rms values of fluctuations in $R_{nl}$,
as a function of the difference in number open of channels in the two voltage probes,
$N_{V+}-N_{V-}$, at fixed values of $N_{V+}+N_{V-}$.
The data in a) presents $\Delta R_{nl}$ values for $N_{V+}+N_{V-}$, that have been averaged
over all possible combinations of $N_{V+}-N_{V-}$.}
\end{figure}

Figure~\ref{fig:voltfluc} presents the results from these numerical
simulations. We focus on analysis of the fluctuations in $R_{nl}$,
since the mean values of $R_{nl}$ simply always gave zero, in
agreement with the experimental results \cite{bardarson2006condmat}.
Figure~\ref{fig:voltfluc}a shows the dependence of the fluctuations
in $R_{nl}$ on the total number of open channels in the voltage
probes $N_{V+}$ + $N_{V-}$, for systems with time-reversal symmetry
($\beta = 1$) and broken time-reversal symmetry ($\beta = 2$). The
results in Fig.~\ref{fig:voltfluc}b show that the fluctuations in
$R_{nl}$ have (besides a strong dependence on $N_{V+}$ + $N_{V-}$ as
in Fig.~\ref{fig:voltfluc}a) a weak dependence on the difference
$N_{V+}$ - $N_{V-}$ (presented only for the case $\beta = 2$). This
is not further studied in detail, and the results in
Fig.~\ref{fig:voltfluc}a present values that are averaged over all
the possible combinations $N_{V+}$ - $N_{V-}$, as in the analysis of
the experimental results.

\begin{figure}[p]
\includegraphics[width=8cm]{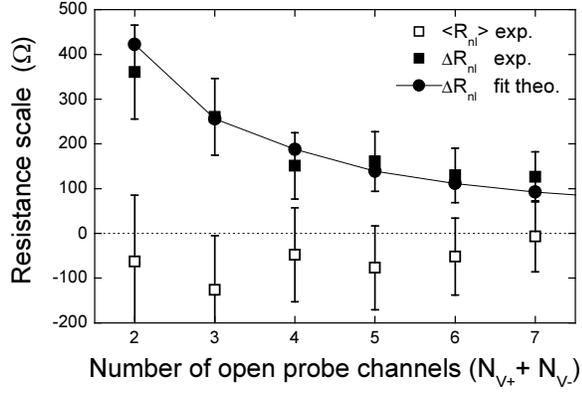}
\caption{\label{fig:flucanalysis} Dependence of the mean $\langle  R_{nl} \rangle $
and rms standard deviation $\Delta R_{nl}$ of
fluctuations in the measured non-local resistance $R_{nl}$, as a function of the
total number open of channels in the voltage probes $N_{V+}+N_{V-}$
(squared dots).
The statistics are from sets of data as in Fig.~\ref{fig:gatedependence},
but with time-reversal symmetry broken by weak magnetic fields ($\beta = 2$).
The round dots with solid line present a fit of the theoretical
model that describes the values of $\Delta R_{nl}$ (see text for details).}
\end{figure}

Qualitatively, the theoretical results of Fig.~\ref{fig:voltfluc}a
agree very well with the experimental results of
Fig.~\ref{fig:flucanalysis}. However, while the model system gives
non-local resistance $R_{nl}$ values that fluctuate with an rms
value of a few k$\Omega$, the experimental values ($\beta$ = 2) are
only of order 200 $\Omega$. The numerical and experimental values
differ by a factor of about 20, as illustrated by the fit in
Fig.~\ref{fig:flucanalysis}: Fitting the theoretical data points of
Fig.~\ref{fig:voltfluc} on the experimental values, using a simple
pre-factor that scales the theoretical values as fitting parameter,
gives $0.05 \approx \frac{1}{20}$ for this pre-factor.

The discrepancy between the numerical and the experimental results
is most likely due to orbital dephasing for electrons inside the
quantum dot. Moreover, such dephasing is possibly consistent with
the simple scaling that was needed to obtain agreement between
theory and experiment. Theory for two-terminal quantum dots gives
that the influence of dephasing on the amplitude of conductance
fluctuations is that it scales the amplitude down by a factor $(1 +
\tau_{d} / \tau_{\phi})$, where $\tau_{d}$ is the mean dwell time in
the dot and $\tau_{\phi}$ is the dephasing time
\cite{baranger1995prb,brouwer1995prb}. Assuming a similar approach
for our system, then gives $(1 + \tau_{d} / \tau_{\phi}) \approx
20$. However, it is at this stage not clear whether this theory work
for two-terminal dots can be applied to our system, and we can also
not rule out that the effective electron temperature of about 400~mK
plays a role in reducing the fluctuation amplitude. Nevertheless, it
is interesting to compare the factor $(1 + \tau_{d} / \tau_{\phi})$
that we need to use here to independently determined values for
$\tau_{d}$ and $\tau_{\phi}$.

For the dwell time we use the expression\cite{marcus1993prb}
$\tau_{d} = h / N_{\Sigma} \Delta_m$, where $N_{\Sigma}$ the total
number of open modes in the contacts to the dot, and $\Delta_m = 2
\pi \hbar^2/m^{*} A_{dot}$ the mean spacing between energy levels in
the dot. This gives $\tau_{d} \approx 470 \; \rm{ps}$ for our system
tuned to have all four QPCs at $N=1$. The most reliable method for
estimating the dephasing time $\tau_{\phi}$ is based on a
measurement of the depth of the weak localization dip in the
two-terminal conductance
\cite{marcus1997chasolfrac,huibers1998Aprl}. In such measurements on
our system (with all four QPCs tuned to $N=1$) we observed a weak
localization dip of $\delta g = 0.045 \pm 0.01 (e^2/h)$ in a
background of $0.88(e^2/h)$ (the large error bar is again due to the
fact that we could only average over a few independent fluctuations
in the two-terminal conductance). Following
Refs.~\onlinecite{marcus1997chasolfrac,huibers1998Aprl}, we derive
the dephasing time using $\delta g = (e^2/h) \cdot N/(2N+N_{\phi})$
and $\tau_{\phi} = h/N_{\phi} \Delta_m$. Here $N_{\phi}$ is the
number of open modes to a fictitious voltage probe that is
responsible for dephasing in the dot, and $N$ is now the number of
modes per lead for a two-terminal dot, so we set it to $N=2$ for our
system with four QPCs tuned to $N=1$. This yields $N_{\phi} \approx
40$ and $\tau_{\phi} = 46 \pm 12 \; \rm{ps}$ for our system at $\sim
400 \; \rm{mK}$, in reasonable agreement with earlier two-terminal
studies on similar systems
\cite{marcus1997chasolfrac,huibers1998Aprl,huibers1999prl}.

The independently determined values for $\tau_{d}$ and $\tau_{\phi}$
give for the factor $(1 + \tau_{d} / \tau_{\phi}) \approx 11$, which
differs only by a factor $\sim 2$ from the value 20 that we obtained
from the scaling. This supports the assumption that the reduction in
the fluctuation amplitude $\Delta R_{nl}$ is due to orbital
dephasing inside the dot. However, the above analysis is only valid
for all QPCs of the dot tuned to $N=1$. While we can apply a single
scaling factor for all values of $N_{V+}$ + $N_{V-}$ in
Fig.~\ref{fig:flucanalysis}, $\tau_{d}$ decreases in fact
significantly with increasing $N_{V+}$ + $N_{V-}$. Moreover, it is
tempting to consider that the reduction in $\Delta R_{nl}$ can be
understood by assuming that dephasing in the dot increases the value
of $N_{V+}$ + $N_{V-}$ to an effective value of $N_{V+}$ + $N_{V-} +
N_{\phi}$. However, this is clearly not in agreement with the
observed drop in $\Delta R_{nl}$ over the interval $N_{V+}$ +
$N_{V-}$ in Fig.~\ref{fig:flucanalysis}. This indicates that new
theoretical work specific for the role of dephasing for a
four-terminal dot is needed for a complete understanding of the data
in Fig.~\ref{fig:flucanalysis}.

\section{\label{sec:conclusions}Conclusions}

We investigated quantum fluctuations in electron transport with a
ballistic, chaotic quantum dot that was strongly coupled to four
reservoirs via quantum point contacts. The four-terminal geometry
allowed for studying fluctuations in the non-local resistance. We
used the dependence of the non-local resistance fluctuations on bias
voltage, gate voltage and magnetic field to show that these are the
equivalent of universal conductance fluctuations in two-terminal
systems, and we showed that with a four-terminal system these
fluctuations can be studied without being hindered by
Coulomb-blockade and weak-localization effects. Furthermore, the
four-terminal geometry was used to demonstrate directly that the
amplitude of fluctuations in electron transport is reduced when the
coupling between a quantum dot system and voltage probes is
enhanced. Here, we obtain good qualitative agreement with a model
based on Landauer-B\"{u}ttiker formalism and random matrix theory,
but a quantitative evaluation indicates that there is an intrinsic
orbital dephasing mechanism that reduces the amplitude of the
non-local resistance fluctuations. Our results are of importance for
further work with four-terminal quantum dots on dephasing and
electron-spin dynamics in such systems, where the electron-transport
signals of interest will always have fluctuations of the type that
is reported here.

\begin{acknowledgments}
We thank Dominik Zumb\"{u}hl and Carlo Beenakker for discussions,
and the Dutch Foundation for Fundamental Research on Matter (FOM)
for financial support. JHB acknowledges support by the European
Community's Marie Curie Research Training Network under contract
MRTN-CT-2003-504574, Fundamentals of Nanoelectronics.
\end{acknowledgments}

\newpage


\begin{thebibliography}{99}

\bibitem{altshuler1998phystod}
B. L. Al'tshuler and P. A. Lee, Physics Today, Issue of December 1988, p. 36.

\bibitem{beenakker1991ssp}
C. W. J. Beenakker and H. van Houten,
Sol. State Phys. \textbf{44}, 1 (1991).

\bibitem{beenakker1997rmp}
C. W. J. Beenakker,
Rev. Mod. Phys. \textbf{69}, 731 (1997).

\bibitem{imry2002book}
Y. Imry,
\textit{Introduction to Mesoscopic Physics} (Oxford University Press, Oxford, 2002).

\bibitem{huibers1998Aprl}
A. G. Huibers, M. Switkes, and C. M. Marcus,
Phys. Rev. Lett. \textbf{81}, 200 (1998).

\bibitem{huibers1998Bprl}
A. G. Huibers, S. R. Patel, C. M. Marcus, P. W. Brouwer, C. I.
Duru\"{o}z, and J. S. Harris, Phys. Rev. Lett. \textbf{81}, 1917
(1998).

\bibitem{alves2002prl}
E. R. P. Alves and C. H. Lewenkopf,
Phys. Rev. Lett. \textbf{88}, 256805 (2002).

\bibitem{zumbuhl2005prb}
D. M. Zumb\"{u}hl, J. B. Miller, C. M. Marcus, D. Goldhaber-Gordon,
J. S. Harris, K. Campman, and A. C. Gossard, Phys. Rev. B
\textbf{72}, 081305 (2005).

\bibitem{buttiker1986prl}
M. B\"{u}ttiker,
Phys. Rev. Lett. \textbf{57}, 1761 (1986).

\bibitem{benoit1987prl}
A. Benoit, C. P. Umbach, R. B. Laibowitz, and R. A. Webb,
Phys. Rev. Lett. \textbf{58}, 2343 (1987).

\bibitem{skocpol1987prl}
W. J. Skocpol, P. M. Mankiewich, R. E. Howard, L. D. Jackel, D. M.
Tennant, and A. D. Stone, Phys. Rev. Lett. \textbf{58}, 2347 (1987).

\bibitem{haucke1990prb}
H. Haucke, S. Washburn, A. D. Benoit, C. P. Umbach, and R. A. Webb,
Phys. Rev. B, \textbf{41}, 12454 (1990).

\bibitem{buttiker1989prb}
M. B\"{u}ttiker, Phys. Rev. B. \textbf{40}, 3409 (1989).

\bibitem{baranger1994prl}
H. U. Baranger and P. A. Mello, Phys. Rev. Lett. \textbf{73}, 142
(1994).

\bibitem{potok2002prl}
R. M. Potok, J. A. Folk, C. M. Marcus, and V. Umansky,
Phys. Rev. Lett. \textbf{89}, 266602 (2002).

\bibitem{folk2007aps}
A. Venkatesan, S. Frolov, J. Folk, and W. Wegscheider, 2007 March
Meeting Bulletin of the American Physical Society (unpublished),
Abstract H12.00012.

\bibitem{jedema2001nature}
F. J. Jedema, A. T. Filip, and B.J. van Wees,
Nature \textbf{410}, 345 (2001).

\bibitem{folk2003science}
J. A. Folk, R. M. Potok, C. M. Marcus, and V. Umansky,
Science \textbf{299}, 679 (2003).

\bibitem{zumbuhl2001aps}
D. M. Zumb\"{u}hl, J. B. Miller, J. A. Folk, S. K. Watson, S. R.
Patel, C. M. Marcus, C. I. Duru\"{o}z, and James S. Harris, 2001
March Meeting Bulletin of the American Physical Society
(unpublished), Abstract C25.006.

\bibitem{beenakker2006prb}
C. W. J. Beenakker,
Phys. Rev. B \textbf{73}, 201304 (2006).

\bibitem{bardarson2006condmat}
J. H. Bardarson, I. Adagideli, and P. Jacquod,
Phys.\ Rev.\ Lett.\ {\bf 98}, 196601 (2007).

\bibitem{landauer1957ibm}
R. Landauer,
IBM J. Res. Dev. \textbf{1}, 223 (1957).

\bibitem{izrailev1990physrep}
F. M. Izrailev,
Phys. Rep. \textbf{196}, 299 (1990).

\bibitem{fyodorov2000jetp}
Y.V. Fyodorov and H.-J. Sommers,
JETP Lett. {\bf 72}, 422 (2000).

\bibitem{vanwees1988prl}
B. J van Wees, H. van Houten, C. W. J. Beenakker, J. G. Williamson,
L. P. Kouwenhoven, D. van der Marel, and C. T. Foxon, Phys. Rev.
Lett. \textbf{60}, 848 (1988).

\bibitem{wharam1988jpc}
D. A. Wharam, T. J. Thornton, R. Newbury, M. Pepper, H. Ahmed, J. E.
F. Frost, D. G. Hasko, D. C. Peacockt, D. A. Ritchie, and G. A. C.
Jones, J. Phys. C: Solid State Phys. \textbf{21}, L209 (1988).

\bibitem{debray1989prl}
P. Debray, J.-L. Pichard, J. Vicente, and P. N. Tung, Phys. Rev.
Lett. \textbf{63}, 2264 (1989).

\bibitem{folk2001prl}
J. A. Folk, S. R. Patel, K. M. Birnbaum, C. M. Marcus, C. I.
Duru\"{o}z, and J. S. Harris, Phys. Rev. Lett. \textbf{86}, 2102
(2001).

\bibitem{zumbuhl2004prb}
D. M. Zumb\"{u}hl, J. B. Miller, C. M. Marcus, V. I. Fal'ko, T.
Jungwirth, and J. S. Harris, Phys. Rev. B \textbf{69}, 121305
(2004).

\bibitem{marcus1997chasolfrac}
C. M. Marcus, S. R. Patel, A. G. Huibers, S. M. Cronenwett, M.
Switkes, I. H. Chan, R. M. Clarke, J. A. Folk, S. F. Godijn, K.
Campman, and A. C. Gossard, Chaos Solitons Fractals \textbf{8}, 1261
(1997); cond-mat/9703038.

\bibitem{huibers1999prl}
A. G. Huibers, J. A. Folk, S. R. Patel, C. M. Marcus, C. I.
Duru\"{o}z, and J. S. Harris, Phys. Rev. Lett. \textbf{83}, 5090
(1999).

\bibitem{marcus1993prb}
C. M. Marcus, R. M. Westervelt, P. F. Hopkins, and A. C. Gossard,
Phys. Rev. B \textbf{48}, 2460 (1993).

\bibitem{chan1995prl}
I. H. Chan, R. M. Clarke, C. M. Marcus, K. Campman, and A. C.
Gossard, Phys. Rev. Lett. \textbf{74}, 3876 (1995).

\bibitem{beenakker1988prb}
C. W. J. Beenakker and H. van Houten,
Phys. Rev. B \textbf{37}, 6544 (1988).

\bibitem{vanhouten1989prb}
H. van Houten, C. W. J. Beenakker, J. G. Williamson, M. E. I.
Broekaart, P. H. M. van Loosdrecht, B. J. van Wees, J. E. Mooij, C.
T. Foxon, and J. J. Harris, Phys. Rev. B \textbf{39}, 8556 (1989).

\bibitem{kobayashi2002jpsjap}
K. Kobayashi, H. Aikawa, S. Katsumoto,  and Y. Iye,
J. Phys. Soc. Japan, \textbf{71}, 2094 (2002).

\bibitem{seelig2003prb}
G. Seelig, S. Pilgram, A.N. Jordan, and M. B{\"u}ttiker,
Phys. Rev. B \textbf{68}, 161310 (2003).

\bibitem{brouwer1995prb}
P. W. Brouwer and C. W. J. Beenakker,
Phys. Rev. B \textbf{51}, 7739 (1995).

\bibitem{brouwer1997prb}
P. W. Brouwer and C. W. J. Beenakker,
Phys. Rev. B \textbf{55}, 4695 (1997).

\bibitem{errorbars}
Analysis of the error bars is in particular relevant for low values
of $N_{V+} + N_{V-}$, for which we could only obtain a few
independent fluctuations. We used that we had the largest amount of
data for traces with $N_{V+}$ + $N_{V-} = 7$. Here we had sufficient
data to estimate $\Delta R_{nl} = 126 \pm 56 \; \Omega$. For traces
with lower $N_{V+}$ + $N_{V-}$, we used that the number of
independent fluctuations $n_{f}$ along a trace with constant
$N_{V+}$ + $N_{V-}$ is simply proportional to $N_{V+}$ + $N_{V-}$
for data sets as in Fig.~\ref{fig:gatedependence}, since it is
proportional to the trace length through the data set. We assume a
gaussian distribution for the fluctuations. Then, we used that for a
finite number $n_{f}$ of independent fluctuations, the error bar for
$ \langle R_{nl} \rangle $ depends on $n_{f}$ as $1/\sqrt{n_{f}}$,
and for $\Delta R_{nl}$ (on the same scale) as $1/\sqrt{2n_{f}}$.

\bibitem{groundingnote}
Note that this definition for ground differs from the grounding used
the in the experiment. However, since in our experiments the values
of the gate voltages were much larger than the values of $V_{bias}$,
our theoretical model is a valid description of the experiment.

\bibitem{bardarson2005prb}
J. H. Bardarson, J. Tworzyd{\l}o, and C. W. J. Beenakker,
Phys. Rev. B \textbf{72}, 235305 (2005).

\bibitem{baranger1995prb}
H. U. Baranger and P. A. Mello,
Phys. Rev. B \textbf{51}, 4703 (1995).

\end{thebibliography}
\end{document}